\documentclass[twocolumn,showpacs,preprintnumbers,amsmath,amssymb,floatfix]{revtex4}
\usepackage{graphicx}
\usepackage{afterpage}
\usepackage{amsmath}
\usepackage{dcolumn}
\usepackage{bm}

\begin{document}

\preprint{APS/123-QED}

\title{Nonlinear magneto-optical resonances at $D_1$ excitation of $^{85}$Rb and $^{87}$Rb in an extremely thin cell}

\author{M.~Auzinsh$^1$}
\email{Marcis.Auzins@lu.lv}
\author{R.~Ferber$^1$}%
\author{F.~Gahbauer$^1$}%
\author{A.~Jarmola$^1$}%
\author{L.~Kalvans$^1$}%
\author{A.~Papoyan$^2$}%
\author{D.~Sarkisyan$^2$}%
\affiliation{ $^1$Laser Centre, The University of Latvia, 19 Rainis
Boulevard, LV-1586 Riga, Latvia\\
$^2$Institute for Physical Research, NAS of Armenia, Ashtarak-0203, Armenia}%


\date{\today}

\begin{abstract}
Nonlinear magneto-optical resonances have been measured in an extremely thin cell (ETC) for the $D_1$ transition of 
rubidium in an atomic vapor of natural isotopic composition. All hyperfine transitions of both isotopes have been 
studied for a wide range of laser power densities, laser detunings, and ETC wall separations. Dark resonances in the 
laser induced fluorescence (LIF) were observed as expected 
when the ground state total angular momentum $F_g$ was greater than or equal to the excited state total angular 
momentum $F_e$. Unlike the case of ordinary cells, the width and contrast of dark resonances formed in the ETC
dramatically depended on the detuning of the laser from the exact atomic transition. 
A theoretical model based on the optical Bloch equations was applied to calculate the shapes of the 
resonance curves. The model, which had been developed previously for ordinary vapor cells, averaged over the contributions from different atomic velocity groups, considered all neighboring 
hyperfine transitions, took into account the splitting and mixing of magnetic sublevels in an external magnetic field, and included a 
detailed treatment of the coherence properties of the laser radiation. Such a theoretical approach had successfully 
described nonlinear magneto-optical resonances in ordinary vapor cells.  However, to describe the resonances in the 
ETC, key parameters such as the ground state relaxation rate, exited state relaxation rate, Doppler width, and Rabi frequency
had to be modified significantly in accordance with the ETC's unique features. The level of agreement between the 
measured and calculated resonance curves achieved for the ETC was similar to what could be accomplished for ordinary cells. 
However, in the case of the ETC, it was necessary to fine-tune parameters such as the background and the Rabi frequency
for different transitions, whereas for the ordinary cells, these parameters were identical for all transitions. 
\end{abstract}
\pacs{32.60.+i,32.80.Xx,32.10.Fn}
\maketitle

\section{\label{Intro:level1}Introduction}
Atomic vapors confined between walls separated by only a few hundred
nanometers have the potential for interesting applications in
magnetometry and optoelectronics~\cite{Sarkisyan:2009,Sargsyan:2008}. 
Spectroscopic cells that have
two walls separated by a distance that is on the order of the
wavelength of visible or near infra-red light are known as extremely
thin cells (ETCs)~\cite{Sarkisyan:2001}. If the laser radiation
propagates in a direction perpendicular to the ETC walls, fluorescence  
will be observed preferentially from those atoms whose velocity vectors
have a small normal component with respect to the walls, since atoms flying
rapidly towards an ETC wall will collide with it before being able
to fluoresce. As a result, sub-Doppler spectroscopic resolution can
be achieved in an ETC \cite{Sarkisyan:2001}. This useful feature
can be exploited to study physical phenomena in atomic systems
with hyperfine manifolds that are not resolved under normal
conditions, such as resonances at zero magnetic field in a
plot of laser induced fluorescence (LIF) versus magnetic field. These
resonances can be dark~\cite{Alzetta:1976,Schmieder:1970} or
bright~\cite{Dancheva:2000}. They arise when at zero magnetic field 
the ground state magnetic sublevels form a quantum superposition state, which is not coupled to the
exciting laser field in the case of dark resonances, but coupled strongly in the
case of bright resonances. Although the theoretical description of these resonances is
straightforward in principle, the theoretical calculations have failed at
times to reproduce accurately the experimental signals, in part, because in most
available systems the excited state hyperfine levels were not resolved
under Doppler broadening. The development of the ETC offered a way to address this
issue, because the ETC makes possible sub-Doppler resolution. This
approach was taken by Andreeva and co-workers~\cite{Andreeva:2007b}
for the case of cesium. However, the results contained some surprises in that some
resonances that would be bright in an ordinary cell appeared dark in
the ETC. 

The goal of the present study was to obtain experimental resonance
signals with an ETC for the rubidium $D_1$ transition with high accuracy 
and as a function of several well-defined parameters. Then, we planned to apply
to these data a theoretical model that had been developed 
to describe nonlinear magneto-optical resonances in ordinary vapor
cells~\cite{Auzinsh:2008,Auzinsh:2009}, to see if we could obtain an adequate 
theoretical description of bright and dark resonances in ETCs with this model.

Nonlinear magneto-optical resonances (see, for example, the review paper by Budker et al.~\cite{Budker:2002})
are closely related to the ground state Hanle effect, 
which was first observed by Lehmann and Cohen-Tannoudji in 1964~\cite{Lehmann:1964}. 
Schmieder~\cite{Schmieder:1970} and later Alzetta~\cite{Alzetta:1976} observed dark resonances in 
alkali atoms. The first observations of coherence resonances by means of lasers rather than spectral lamps
were carried out by Ducloy et al.~\cite{Ducloy:1973} in fluorescence and by Gawlik et al.~\cite{Gawlik:1974} 
in connection with the non-linear Faraday effect.
In 1978, Piqu\'e successfully applied a theoretical model based on the optical Bloch equations 
to describe  measurements of dark resonances in the $F_g=2\rightarrow F_e=1$ transition of the $D_1$ line in 
a beam of sodium atoms~\cite{Picque:1978}.
In 2000, Dancheva et al.~\cite{Dancheva:2000} observed bright resonances for the first time in the $D_1$ and $D_2$ 
transitions of rubidium in an atomic vapor. A theoretical model~\cite{Blushs:2004} 
based on the optical Bloch equations was 
applied to the experimental study of bright and dark resonances in the $D_2$ transition of cesium observed in
a vapor cell.~\cite{Papoyan:2002}.  However, the calculations predicted a bright resonance 
for the $F_g=4\rightarrow F_e=3,4,5$ transition, whereas for linearly polarized excitation it appeared to 
be dark in the experiment. Andreeva et al.~\cite{Andreeva:2007b} studied bright and dark resonances in 
ETCs and suggested that spin-changing collisions of the atoms with the walls could change some bright resonances 
into dark resonances.

Our initial approach~\cite{Auzinsh:2008,Auzinsh:2009} to the difficulties in describing bright and
dark resonances in atomic vapors was to develop further a theoretical 
model~\cite{Blushs:2004} that would take into account the Doppler effect. 
Besides averaging
over the Doppler profile, the model also accounted for all nearby
hyperfine transitions, the splitting and mixing of the magnetic
sublevels in the magnetic field, and the coherence properties of the
laser radiation. 
We applied such a model first to the cesium $D_1$
transition~\cite{Auzinsh:2008}, which is almost resolved under
Doppler broadening, and then to the rubidium $D_1$
transition~\cite{Auzinsh:2009}, which is partially resolved. The
model was successful in describing experimentally measured
resonances in ordinary cells for a wide variety of laser power
densities, beam diameters, and laser frequency detunings. It required only a
few parameters to be adjusted. These parameters were the conversion
between the Rabi frequency and the laser power density, the conversion
between the laser beam diameter and the ground state relaxation time, and the
percentage of background from scattered LIF. Once the parameters 
were determined, the same values were used for all data sets. 
This model served as the starting point of our efforts in the present
study to describe bright and dark resonances in the ETC.

\begin{figure}[htbp]
    \centering
        \resizebox{\columnwidth}{!}{\includegraphics{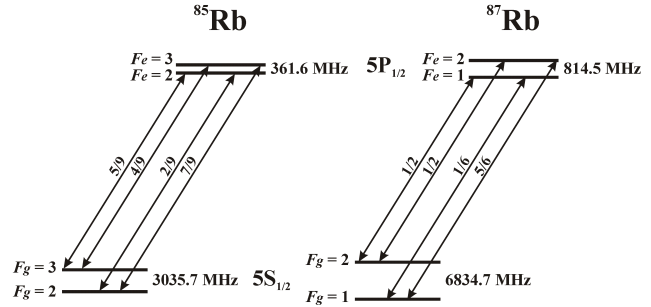}}
    \caption{\label{fig:levels} Hyperfine level structure and transitions of the  $D_1$ line of rubidium. 
The fractions on the arrows indicate the relative transition strengths~\cite{Steck:rubidium85,Steck:rubidium87}.}
\end{figure}

\begin{figure}[htbp]
    \centering
        \resizebox{\columnwidth}{!}{\includegraphics{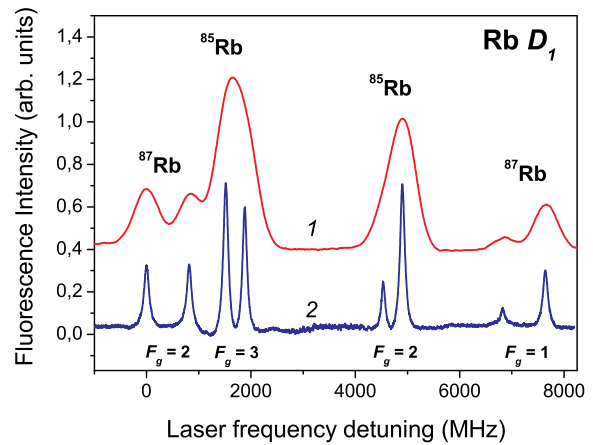}}
    \caption{\label{fig:LIF_excitation} (color online) LIF excitation spectra of the Rb $D_1$ line. Curve 1: in an ordinary cell at 
10 mW/cm$^2$ excitation; 
curve 2: in an ETC at 10 mW/cm$^2$ excitation and wall separation $L=\lambda/2$, where $\lambda$ is
the wavelength of the laser radiation.}
\end{figure}

The level scheme of the $D_1$ line in
$^{85}$Rb and $^{87}$Rb is shown in Fig.~\ref{fig:levels}. 
The allowed transitions between the ground
and excited states are represented by the arrows with the relative
transition strengths given by the fractions next to the arrows. The
energy difference between the excited state hyperfine levels is
361.6 MHz for $^{85}$Rb~\cite{Steck:rubidium85} and 814.5 MHz for $^{87}$Rb~\cite{Steck:rubidium87}. 
In an ordinary vapor cell, the full width at half maximum (FWHM) of the Doppler
broadening for the rubidium $D_1$ transition at room temperature would be about 500 MHz.
In the ETC, however, the Doppler broadening is reduced, because
atoms flying parallel to the walls are more likely to be observed
in the experiment. The LIF excitation spectra of the $D_1$ line of natural rubidium
for both an ordinary vapor cell and an ETC are shown in 
Fig.~\ref{fig:LIF_excitation}. In the ordinary cell, the
two excited state hyperfine levels of  $^{87}$Rb are almost resolved
under Doppler broadening, but the hyperfine levels of  $^{85}$Rb are
practically unresolved. In the ETC, however, each transition can be
distinguished easily.

\section{\label{Experiment:level1}Experiment}
\begin{figure}[htbp]
    \centering
        \resizebox{0.5\columnwidth}{!}{\includegraphics{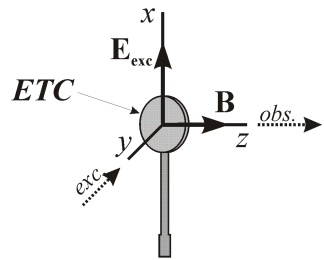}}
    \caption{\label{fig:geometry} Experimental geometry. The relative orientation of the laser beam (exc), 
laser light polarization (\textbf{E$_{exc}$}), magnetic field (\textbf{B}), observation direction (obs), 
and ETC walls are shown.
}
\end{figure}

The ETC was produced at the Institute of Physical Research in
Ashtarak, Armenia, while the experiments described here were performed in Riga. 
The principle of the ETC is described in~\cite{Sarkisyan:2001, Sarkisyan:2009}. 
“Basically, it consists of two perfectly polished window wafers made from 
YAG crystal that are glued in such a way that their internal separation 
varies from 50 nm to about 1.8 $\mu$m. Rubidium is stored in a sapphire tube 
bonded to the YAG windows and terminated by a small glass tube extension 
which makes it possible to fill the cell with rubidium and seal it.
The cell was operated in a two-chambered
oven, which keeps the YAG windows at about 200$^o$C and the
rubidium arm at about 150$^o$C.
The laser was a DL 100 external cavity single mode diode laser with a 
wavelength of 794.3 nm and a typical linewidth of a few megahertz, 
which was produced by Toptica, A.~G., of Graefelfing, Germany. 
The laser beam passed through a Glan-Thompson polarizer before entering the cell.
The cell was placed at the center of a three-axis Helmholtz coil
system, which rested on a nonmagnetic optical table. 
Two pairs of coils compensated the laboratory magnetic
field, while the third pair of coils was used to scan the magnetic
field in the observation direction. The current in this third pair
of coils was scanned with a Kepco BOP-50-8M bipolar power supply,
which was controlled by an analog signal from a computer.
Figure~\ref{fig:geometry} depicts the relative orientation of the
laser beam (\textit{exc}), the laser radiation's linear polarization vector
(\textbf{E$_{exc}$}), the magnetic field (\textbf{B}), the
observation direction (\textit{obs}), and the ETC walls. The experimental
setup and auxiliary equipment were similar to what was used
in~\cite{Auzinsh:2008}.
Data was acquired with a National Instruments 6024E data acquisition
card. 

    The LIF was detected by means of a Thorlabs 
FDS-100 photodiode as a function of magnetic field for all hyperfine 
transitions of the two isotopes of rubidium (see Fig.~\ref{fig:levels}). 
Fluorescence emerged through the side walls of the ETC.
No polarizers were used in the LIF observation. Signals 
were obtained for various laser power densities between 10 mW/cm$^2$ and 
2000 mW/cm$^2$. The laser beam diameter was usually 0.44 mm (FWHM of intensity), as 
measured by a Thorlabs BP 104-VIS beam profiler, but different 
laser beam diameters were also applied to the cell.

To record the signal, the laser's frequency was maintained at the
value that gave the maximum fluorescence signal for a given transition
at zero magnetic field. The fluorescence was recorded while the
magnetic field was scanned several times, and the scans were
averaged. The laser was not actively stabilized, but its frequency
was monitored with a High Finesse WS-7 wavemeter. 
In order to study the dependence of resonance shapes on laser detuning, 
for some observations we used a double scanning technique~\cite{Andreeva:2002}, 
in which the laser frequency was scanned slowly across a transition
while the magnetic field was scanned more rapidly from negative to 
positive values with a triangular waveform. The laser
frequency changed by about 2--5 MHz per second. The typical laser
frequency scan lasted about 1--2 minutes, whereas the typical period
for the magnetic field scan was 1 second. In this manner, a series
of resonance signals at laser frequencies differing by 2--5 MHz
could be obtained.

    In addition to the signal, each measurement included a certain amount 
of background. Background from scattered laser light was determined for 
each measurement by tuning the laser off resonance. However, there was an 
additional background associated with scattered LIF, which could not be 
readily identified. Studies in the ordinary cell had suggested that this 
background accounts for a fixed percentage of the signal for a given vapor 
cell. However, it is also possible that the background differs according to 
which isotope is studied, since the total amount of fluorescence changes. 
Furthermore, the background from scattered LIF will likely depend on the thickness of
the cell. Its value can be high because
fluorescence observed by the detector can be reflected 
by the walls of the ETC or scattered while passing through the glue
that held the walls together (see Fig.~\ref{fig:geometry}).
The value of this background was one of the parameters that had to be 
adjusted to find the best fit between experiment and theory.

\section{\label{Theory:level1}Theory}
A theoretical model had been developed previously in order to describe bright and
dark resonances in ordinary vapor cells~\cite{Auzinsh:2008,Auzinsh:2009}, and a detailed 
description of this theoretical model can be found in these references. We summarize it here
briefly.
The model describes the internal atomic dynamics by a semi-classical atomic density matrix $\rho$, 
which depends parametrically on the classical coordinates of the atomic center of mass. 
The time evolution of the density matrix $\rho$ follows the optical Bloch equations (OBEs)~\cite{Stenholm:2005}:
\begin{equation}
        i\hbar \frac{\partial \rho}{\partial t} = \left[\hat{H},\rho \right] + i \hbar\hat{R}\rho.
        \label{obe}
\end{equation}
The relaxation operator $\hat{R}$ includes the spontaneous emission rate, which equals the natural transition 
linewidth $\Gamma_N$, and the ground state relaxation rate $\gamma_g$, which in the case of the ETC corresponds to the 
rate of collisions with the cell wall. We also introduced an exccited state relaxation rate $\gamma_e$ related mostly to 
wall interactions in the ETC, which was negligible in the ordinary cell. 
The Hamiltonian $\hat{H}$ is given by  $\hat{H} = \hat{H}_0 + \hat{H}_B + \hat{V}$.
$\hat{H}_0$ is the unperturbed atomic Hamiltonian and depends on the internal atomic coordinates, 
$\hat{H}_B$ is the Hamiltonian of the atomic interaction with the magnetic field, 
and $\hat{V} = -\hat{\textbf{d}} \cdot \textbf{E}(t)$ is the dipole interaction operator.

Taking into account the classical motion of the atoms, the resulting Doppler shifts, and 
the coherent properties of the laser radiation, 
we apply the rotating wave approximation~\cite{Allen:1975} to the OBEs. This approximation 
yields stochastic differential equations, which can be further simplified by using the 
decorrelation approach~\cite{Kampen:1976}. Since the experimentally observed light 
intensity is averaged over time intervals that are large compared to the characteristic 
time of the phase fluctuations, it is permissible to average over the stochastic differential 
equations following the method described in~\cite{Blushs:2004} and references therein.

After adiabatically eliminating the density matrix elements that correspond to optical coherences, we arrive 
at the equations for the Zeeman coherences:
\begin{align}
\dfrac{\partial\rho_{g_ig_j}}{\partial t} =&\bigl(\Gamma_{p,g_ie_m} + \Gamma_{p,e_kg_j}^{\ast}\bigr)\underset{e_k, e_m
}{ \sum }\bigl(d_1^{g_ie_k}\bigr)^{\ast}d_1^{e_mg_j}\rho_{e_ke_m} \nonumber \\
&- \underset{e_k,g_m}{\sum }\Bigl[\Gamma_{p,e_kg_j}^{\ast} \bigl(d_1^{g_ie_k}\bigr)^{\ast}d_1^{e_kg_m}\rho_{g_mg_j} \nonumber \\
&+\Gamma_{p,g_ie_k} \bigl(d_1^{g_me_k}\bigr)^{\ast}d_1^{e_kg_j}\rho_{g_ig_m}\Bigr] - i\omega_{g_ig_j}\rho_{g_ig_j} \nonumber \\
&+ \underset{e_i, e_j}{\sum}\Gamma_{g_ig_j}^{e_ie_j}\rho_{e_ie_j} -\gamma_g\rho_{g_ig_j} + \lambda\delta\bigl(g_i, g_j\bigr)
\label{rate1}
\end{align}
and
\begin{align}
\dfrac{\partial\rho_{e_ie_j}}{\partial t} =&\bigl(\Gamma_{p,e_ig_m}^{\ast} + \Gamma_{p,g_ke_j}\bigr) \underset{g_k, g_
m}{\sum }d_1^{e_ig_k}\bigl(d_1^{g_me_j}\bigr)^{\ast}\rho_{g_kg_m} \nonumber \\
&- \underset{g_k,e_m}{\sum }\Bigl[\Gamma_{p,g_ke_j} d_1^{e_ig_k}\bigl(d_1^{g_ke_m}\bigr)^{\ast}\rho_{e_me_j} \nonumber \\
&+\Gamma_{p,e_ig_k}^{\ast} d_1^{e_mg_k}\bigl(d_1^{g_ke_j}\bigr)^{\ast}\rho_{e_ie_m}\Bigr]  \nonumber \\
&- i\omega_{e_ie_j}\rho_{e_ie_j} - \Gamma\rho_{e_ie_j}
\label{rate2}.
\end{align}
In these equations $\rho_{g_ig_j}$ and $\rho_{e_ie_j}$ are the density matrix
elements for the ground and excited states, respectively. The first
term in (\ref{rate1}) describes the re-population of the ground
state and the creation of Zeeman coherences from induced
transitions; $\Gamma_{p,g_ie_j}$ and $\Gamma_{p,e_ig_j}^{\ast}$
are the couplings between the laser field and the ground and excited states;
and $d_1^{e_ig_j}$ is the dipole transition matrix element between 
the ground state $i$ and the excited state $j$. The
second term contains the changes of the ground state Zeeman sublevel
population and the creation of ground state Zeeman coherences after
light absorption. The third term describes the destruction of the
ground state Zeeman coherences by the external magnetic field;
$\omega_{g_ig_j}$ is the splitting of the ground state Zeeman
sublevels. The fourth term describes the re-population and transfer
of excited state coherences to the ground state as a result of spontaneous
transitions. We assume our transition to be closed (within the
fine structure, all atoms undergoing spontaneous transitions
return to the initial fine structure state $5S_{1/2}$), and so
$\sum_{e_i, e_j}\Gamma_{g_ig_j}^{e_ie_j} = \Gamma_N$, the natural
linewidth.
The fifth and sixth terms show the relaxation and re-population of the
ground state due to non-optical reasons; in our case it is assumed
to be solely ground state relaxation due to wall collisions. 
We assumed that the atomic equilibrium density outside the interaction 
region is normalized to 1, and so $\lambda = \gamma_g$.

In  equation (\ref{rate2}) the first term describes the light absorbing transitions; 
the second term denotes induced transitions to the ground state; 
the third describes the destruction of ground state Zeeman coherences in the external 
magnetic field, where $\omega_{e_ie_j}$ is the splitting of the excited state Zeeman sublevels; 
and the fourth term denotes the rate of spontaneous decay of the excited state. The excited 
state decay rate is given by $\Gamma=\Gamma_N+\gamma_e$, where $\Gamma_N$ is the natural
linewidth, and $\gamma_e$ represents the 
relaxation rate of the excited state, which results mostly from interactions with the ETC walls. 

The interaction strength $\Gamma_{p,g_ie_j}$ is given by
\begin{equation}
\Gamma_{p,g_ie_j} = \frac{\Omega_R^2}{\left[\left(\frac{\Gamma_N}{2} + \frac{\Delta\omega}{2}\right) \pm i\left(\overline{\omega} - \boldsymbol{k}_{\overline{\omega}}\boldsymbol{v} - \omega_{e_jg_i}\right)\right]}
\label{gammaP1},
\end{equation}
where $\overline{\omega}$ is the central laser frequency, $\Delta\omega$ is the laser line width, and  
the squared Rabi frequency $\Omega_R^2$ is proportional to the laser power density.

The magnetic field not only splits the Zeeman sublevels by an amount
$\omega_{ij}$, but also changes the transition dipole elements by mixing
the magnetic sublevels. The Breit-Rabi
formula~\cite{Breit:1931, Aleksandrov:1993} gives the mixing in the case 
of two hyperfine levels. Although the ground state
hyperfine levels are separated by several gigahertz, all four hyperfine components were taken
into account simultaneously to model the experimental signals. As can be
seen from equation (\ref{gammaP1}), absorption is possible from both hyperfine levels
in one experiment when taking into account the natural
linewidth of the optical transition, the Doppler lineshift, and the laser
detuning from the exact resonance.

The experiments are assumed to take place under stationary excitation conditions 
so that $\partial\rho_{g_ig_j}/\partial t = \partial\rho_{e_ie_j}/\partial t = 0$. 
Thus, one can reduce the differential equations (\ref{rate1}) + (\ref{rate2}) 
to a system of linear equations and solve it to obtain the density matrices 
for the atomic ground and excited states. From the density matrices, one obtains 
the observed fluorescence intensity as follows:
\begin{equation}
        I_f(\tilde{\textbf{e}})=\tilde{I}_0\underset{g_i, e_i, e_j}{\sum}d^{(ob)*}_{g_ie_j}d^{(ob)}_{g_ie_i}\rho_{e_ie_j},
        \label{fluorescence}
\end{equation}
where $\tilde{I}_0$ is a constant of proportionality.

The signal is integrated over the different atomic velocity groups and summed over
the two orthogonal polarization components of the fluorescence and
over all nearby hyperfine transitions, including both ground state
hyperfine levels.

\begin{figure*}[htbp]
    \centering
        \resizebox{\textwidth}{!}{\includegraphics{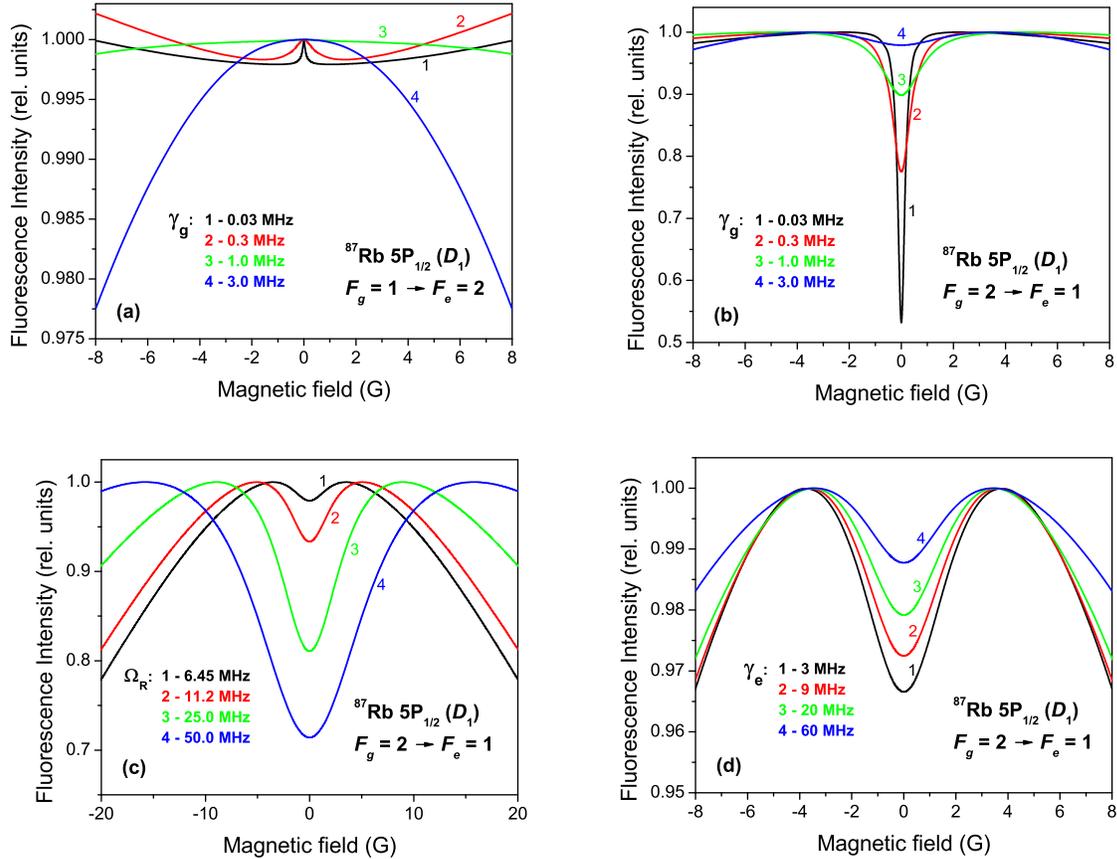}}
        \caption{\label{fig:parameters} (color online)  Theoretical predictions of 
resonance shapes for various values of parameters found in the theoretical model. 
(a) Bright resonance at the $F_g=1\rightarrow F_g=2$ transition of $^{87}$Rb for various 
values of the ground state relaxation rate $\gamma_g$. (b) Dark resonance at the 
$F_g=2\rightarrow F_g=1$ transition of $^{87}$Rb for different values of the ground state
relaxation rate $\gamma_g$. (c) Dark resonance at the $F_g=2\rightarrow F_g=1$ transition 
of $^{87}$Rb for different values of the Rabi frequency $\Omega_R$. (d) Dark resonance at 
the $F_g=2\rightarrow F_g=1$ transition of $^{87}$Rb for different values of the excited 
state relaxation rate $\gamma_e$.}

\end{figure*}

The above model initially was developed for vapor cells of ordinary dimensions. However, 
the geometry of the ETC is quite different from an ordinary cell. In particular, 
the atoms are confined between very narrow walls, which could affect the dynamics 
of the excitation and relaxation processes. For example, in the ordinary cell, 
the ground state relaxation rate $\gamma_g$ is related to the beam diameter and is usually 
on the order of a few tens of kilohertz. In the ETC, however, atoms will collide 
with the wall of the ETC much more frequently. Taking into account the Doppler width 
estimated from Fig.~\ref{fig:LIF_excitation} and the wall separation, one can expect 
ground state relaxation rates on the order of several megahertz in the ETC. 

Figure~\ref{fig:parameters} shows the results for two 
transitions of $^{87}$Rb of theoretical calculations that were performed to 
study the impact on the resonance shapes of various parameters that might depend on
wall thickness in the ETC. Except as otherwise indicated in the figures, the Rabi
frequency used in the calculations was $\Omega_R=6.45$ MHz, and the ground state relaxation
rate was $\gamma_g=3$ MHz. The excited state relaxation rate was $\gamma_e=20$ MHz, 
except in Fig.~\ref{fig:parameters}(a), where it was close to zero for $\gamma_g=0.03$ MHz and
increased to $\gamma_e=20$ MHz for $\gamma_g=3.0$ MHz. 
The Doppler width was
assumed to be 500 MHz for $\gamma_g < 0.3$ and after that decreased linearly to 60
MHz at $\gamma_g=3.0$ MHz. Larger ground and excited state relaxation rates were assumed to be related to 
smaller wall separation, which favors excitation and fluorescence of atoms that have a small
velocity component perpendicular to the walls. As the distance between the ETC
walls decreases, the more likely it becomes for atoms to collide 
with the walls during the atom-light interaction, and so the relaxation
rates increase. 
  
Figures~\ref{fig:parameters}(a) and (b) show effects of changing the ground and excited 
state relaxation rates in the model for the case of a bright and a dark resonance, respectively. 
A ground state relaxation rate $\gamma_g=0.03$ MHz is typical of an ordinary cell, in which
$\gamma_e$ is negligible. In an ETC $\gamma_g$ on the order of several megahertz and $\gamma_e$ 
on the order of twenty megahertz could be realistic values when taking into account frequent wall collisions. 
At slow ground state 
relaxation rates, one observes narrow resonances in both cases, although in the case of the 
bright resonance the contrast is small. As the ground state relaxation rate increases, 
the narrow bright resonance at first gets broader and eventually disappears completely, 
while simultaneously the large-field fluorescence decreases. For the dark resonance, 
the resonance width increases with increasing ground state relaxation rate, while the 
resonance contrast decreases. The large-field fluorescence likewise decreases with 
increasing ground state relaxation rate. Figure~\ref{fig:parameters}(c) shows the dark
resonance for various values of the Rabi frequency $\Omega_R$ for $\gamma_g=3.0$ MHz. 
Figure~\ref{fig:parameters}(c) shows
that the resonance contrast, resonance width, and the value at which the fluorescence
reaches a maximum all increase with increasing Rabi frequency.  
Finally, Figure~\ref{fig:parameters}(d) shows the effect on the resonance shapes of varying
the excited state relaxation rate $\gamma_e$. The excited state
decay rate may also increase as the wall separation decreases. It can be seen that as the
excited state decay rate increases, the resonance contrast decreases, and fluorescence decreases 
less steeply at large magnetic field values. 

The evolution of these calculated signals qualitatively agrees with the evolution of experimentally
measured signals as one goes from ordinary cells with large laser beam diameters to smaller laser 
beam diameters (see~\cite{Auzinsh:2009}, Fig.~9), and finally to an ETC, 
as will become apparent in the next section. We therefore put forth the hypothesis that the
resonance shapes in the ETC can be modelled by the same theoretical model that is used in vapor cells
of ordinary dimensions, but that it is necessary to adjust the values of the ground state relaxation
rate, excited state relaxation rate, Doppler width, and Rabi frequency in accordinance with the
small distance between the ETC cell walls.  
In order to test this hypothesis, we tried to find consistent 
relationships among the cell wall separation and laser power density on the one hand, and the 
ground state relaxation rate, excited state relaxation rate, and Rabi frequency on the other hand for
a large set of measurements performed on all transitions of the $D_1$ line for various laser
power densities and wall separations.       
Among the adjustable parameters of the model are the conversion between 
laser power density $I$ and Rabi frequency $\Omega_R$, the conversion between 
the ETC wall separation and the ground state relaxation rate $\gamma_g$ and excited 
state relaxation rate $\gamma_e$, and the Doppler width, which was 
different in the ETC from the case of the ordinary cell.
The atomic constants used in the model are 
taken from the compilations by Steck~\cite{Steck:rubidium85,Steck:rubidium87}.

\section{\label{Results:level1}Results and Discussion}
\begin{figure*}[htbp]
    \centering
        \resizebox{\textwidth}{!}{\includegraphics{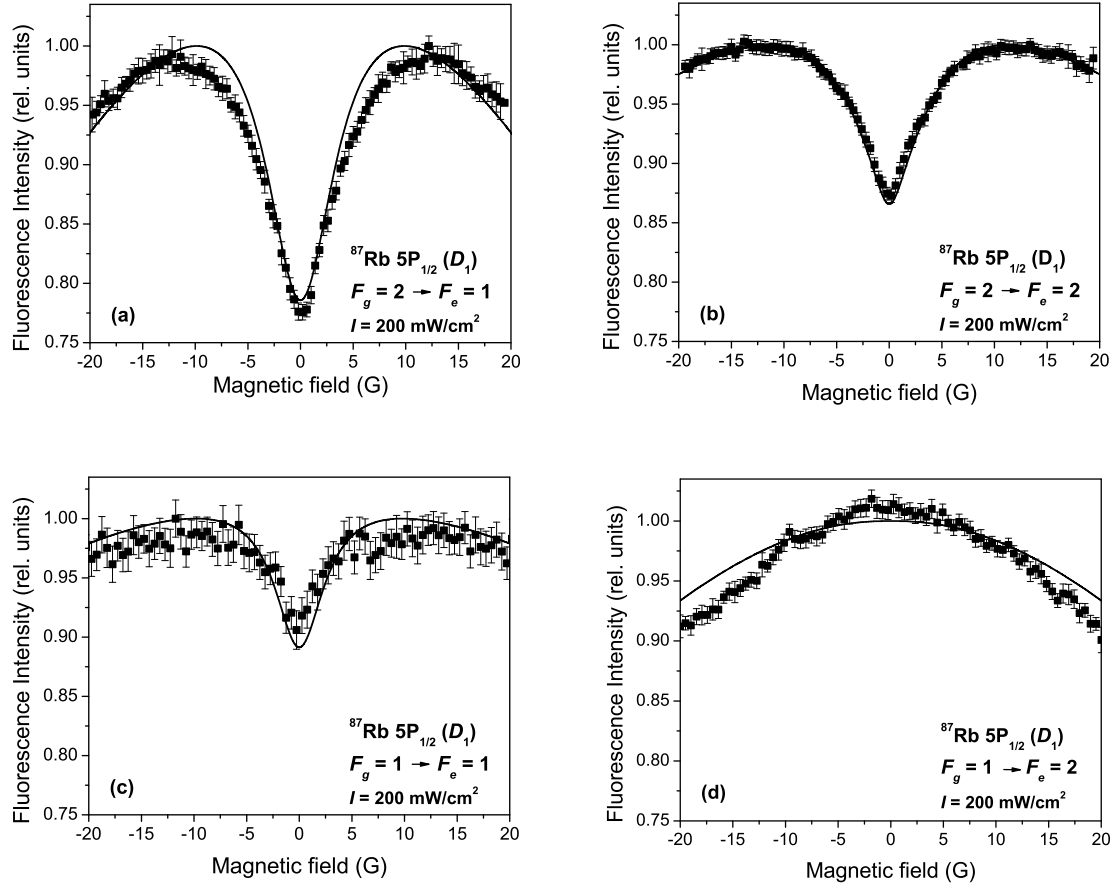}}
    \caption{\label{fig:resonances87} Fluorescence intensity versus magnetic 
field for $^{87}$Rb at $D_1$ excitation. Filled squares, experiment; solid line, theory. 
The excited state, total angular momentum of the ground $F_g$ and excited states $F_e$ 
of the transition and laser power density $I$ are given in each panel.
ETC wall separation $L$ is equal to $\lambda$, the wavelength of the light. 
}
\end{figure*}

\begin{figure*}[htbp]
    \centering
        \resizebox{\textwidth}{!}{\includegraphics{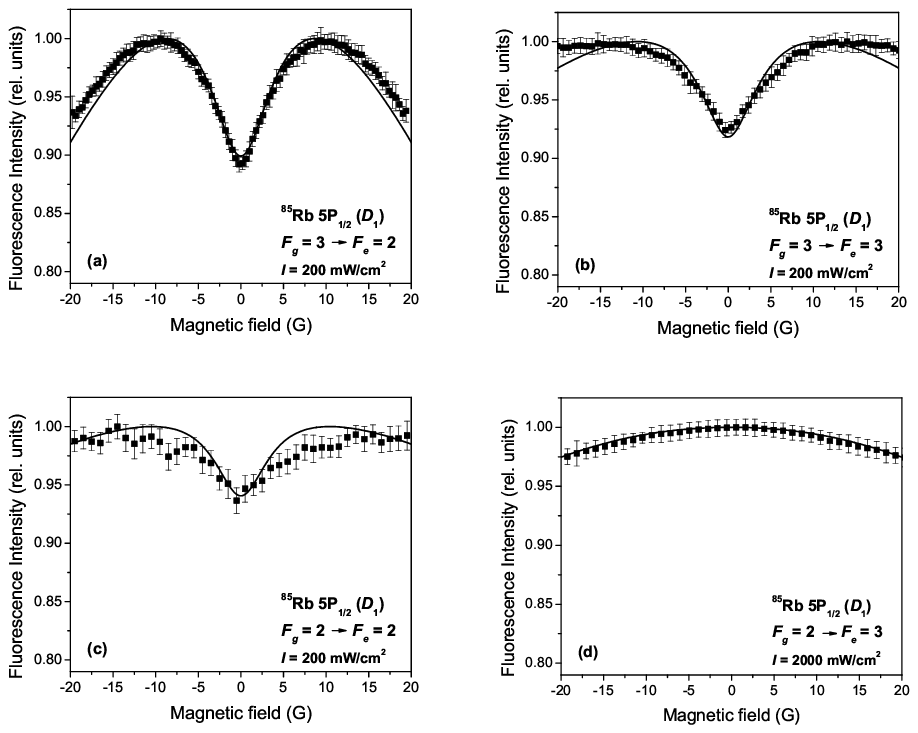}}
    \caption{\label{fig:resonances85} Fluorescence intensity versus magnetic field 
for $^{85}$Rb at $D_1$ excitation. Filled squares, experiment; solid line, theory. 
The excited state, total angular momentum of the ground $F_g$ and excited states 
$F_e$ of the transition, and laser power density $I$ are given in each panel.
$L=\lambda.$
}
\end{figure*}

In general, one expects to observe dark resonances if the total
ground state angular momentum $F_g$ is greater than or equal to the total
excited state momentum $F_e$ and a bright resonance if 
$F_g<F_e$~\cite{Kazantsev:1984,Renzoni:2001a,Alnis:2001}. 
Figures~\ref{fig:resonances87} and~\ref{fig:resonances85} show 
magneto-optical resonances for the
various hyperfine transitions at the $D_1$ line of $^{87}$Rb
and $^{85}$Rb, respectively. Markers represent experimental 
data while the lines are the result of a theoretical calculation. 
The following parameter values were assumed in the theoretical model to describe
all transitions at cell wall separation $L=\lambda$: 
the ground state relaxation rate $\gamma_g$ was 2.88 MHz, the Doppler width was 60 MHz (FWHM), 
and the excited state relaxation rate $\gamma_e$ was 20 MHz.  
Furthermore, for all the transitions of $^{87}$Rb the relationship between Rabi
frequency $\Omega_R$ in Megahertz and laser power density $I$ in mW/cm$^2$ 
was assumed to be $I=0.25$ mW cm$^{-2}$MHz$^{-2}\Omega_R^2$ and a background, discussed in 
Section~\ref{Experiment:level1}, from laser induced
fluorescence of 50\% was assumed. For $^{85}$Rb, the relationship 
$I=0.8$ mW cm$^{-2}$MHz$^{-2}\Omega_R^2$ was assumed for the transitions with equal total angular
momentum $F$ in the ground and excited states and $I=0.4$ mW cm$^{-2}$MHz$^{-2}\Omega_R^2$ for the 
transitions in which $F$ changed. The background from laser
induced fluorescence was assumed to be 37\% for all transitions of $^{85}$Rb. 

Dark resonances were clearly seen as expected at the
$F_g=2\rightarrow F_e=1$, $F_g=2\rightarrow F_e=2$, and
$F_g=1\rightarrow F_e=1$ transitions shown in Fig.~\ref{fig:resonances87}, 
and at the $F_g=3\rightarrow F_e=2$, $F_g=3\rightarrow F_e=3$, and
$F_g=2\rightarrow F_e=2$ transitions shown in Fig.~\ref{fig:resonances85}, 
since $F_g\geq F_e$. However, neither the 
$F_g=1\rightarrow F_e=2$ transition in $^{87}$Rb nor the 
$F_g=2\rightarrow F_e=3$ transition in $^{85}$Rb
gave any indication of
either a bright or dark resonance associated with ground state
coherences, even though one might have expected a bright resonance here
since $F_g<F_e$. This result is entirely consistent with the predictions
of the theoretical model shown in Fig.~\ref{fig:parameters}(a). 
The decrease of the observed fluorescence with
increasing magnetic field was related to a loss of excited state coherence as well
as to the fact that at high magnetic fields the energy separation of individual
transitions between the Zeeman sublevels exceeded the laser linewidth, and, hence,
resonant absorption was reduced. 

\begin{figure*}[htbp]
    \centering
        \resizebox{\textwidth}{!}{\includegraphics{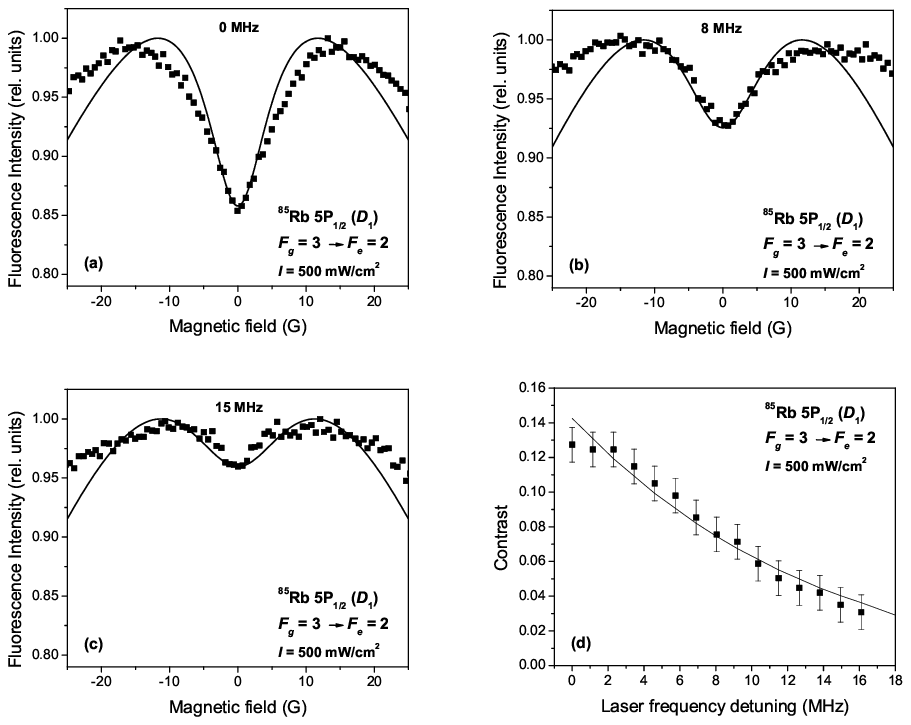}}
    \caption{\label{fig:detuning} (a-c) Fluorescence intensity versus magnetic 
field for the $F_g=3\rightarrow F_e=2$ transition of $^{85}$Rb at $D_1$ excitation at various values 
of the laser detuning with respect to the exact position of the 
$F_g=3\rightarrow F_e=2$ transition; (d) resonance contrast versus laser frequency detuning. 
Filled squares, experiment; solid line, theory.
$L=\lambda$.}
\end{figure*}

It has been shown previously in an ordinary rubidium vapor cell that 
the shape of a nonlinear magneto-optical resonance is sensitive to detuning~\cite{Auzinsh:2009}.
The ETC should be much more sensitive to detuning because of its sub-Doppler characteristics. 
Figure~\ref{fig:detuning} depicts the $F_g=3\rightarrow F_e=2$ transition of $^{85}$Rb for 
various values of the detuning from the point of maximum fluorescence between 0 and 16 MHz. 
The theoretical model takes into account that the ground state relaxation rate $\gamma_g$ increases 
linearly with the detuning $\Delta$
as different velocity groups in the Doppler profile are selected following the relation
$\gamma_g=\gamma_0+0.3\Delta$, with $\gamma_0=2.88$ MHz.
As in the case of the ordinary cell, the contrast decreased as the laser was detuned from 
the resonance, but in the ETC the effect was much more dramatic. In the ETC, a detuning of only
20 MHz resulted in a reduction in the contrast of the dark resonance that was comparable to the reduction
observed in the ordinary cell for a detuning of 300 MHz~\cite{Auzinsh:2009} at the 
$F_g=3\rightarrow F_e=2$ transition of $^{85}$Rb. The theoretically calculated curve does not 
describe the large-field behavior particularly well,
but it does describe very well the change in contrast as a function of detuning, as can be seen
especially in Fig.~\ref{fig:detuning}(d).

\begin{figure*}[htbp]
    \centering
        \resizebox{\textwidth}{!}{\includegraphics{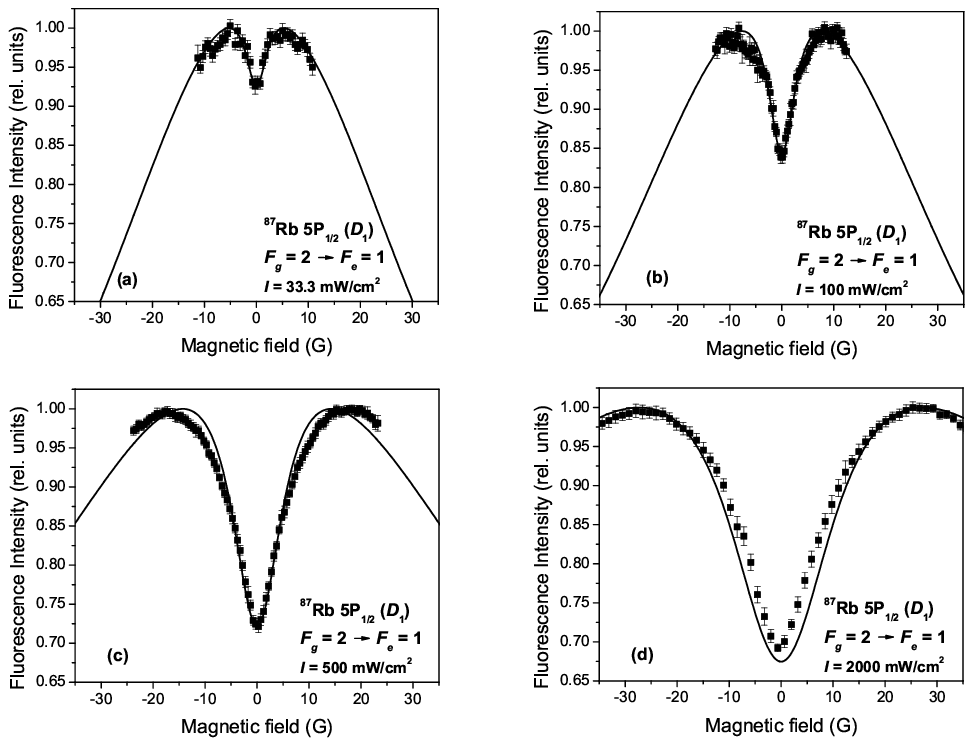}}
    \caption{\label{fig:rb87_11_rabi} Resonance signals for $^{87}$Rb at 
the $F_g=2\rightarrow F_e=1$ transition for different laser power densities $I$. 
Filled squares, experiment; solid line, theory. $L=\lambda$.}
\end{figure*}

\begin{figure*}[htbp]
    \centering
       \resizebox{\columnwidth}{!}{\includegraphics{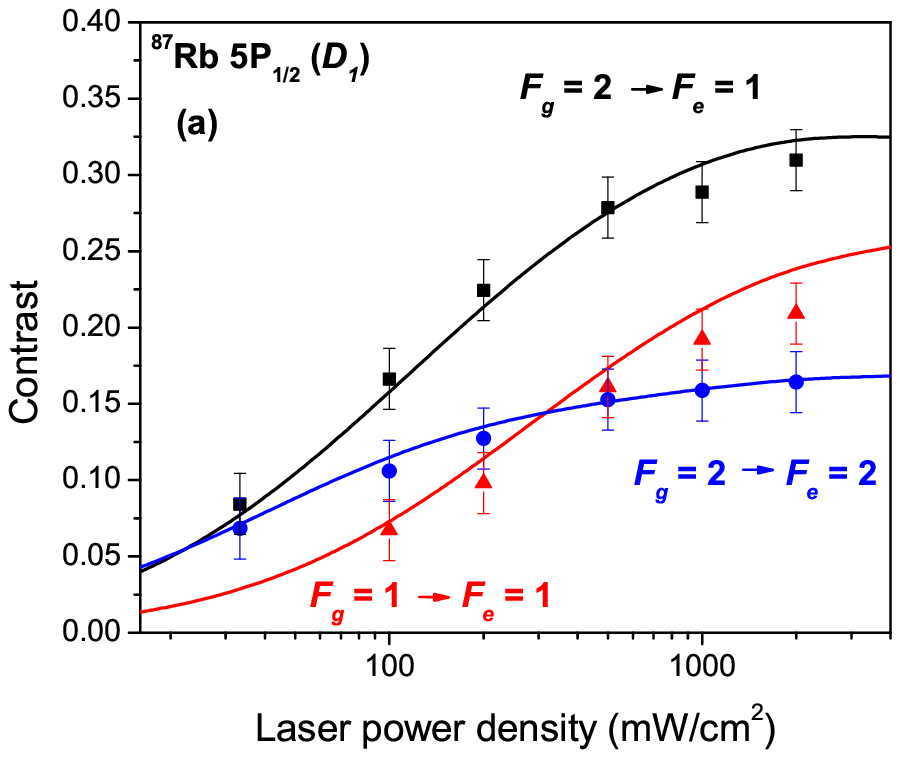}}
       \resizebox{\columnwidth}{!}{\includegraphics{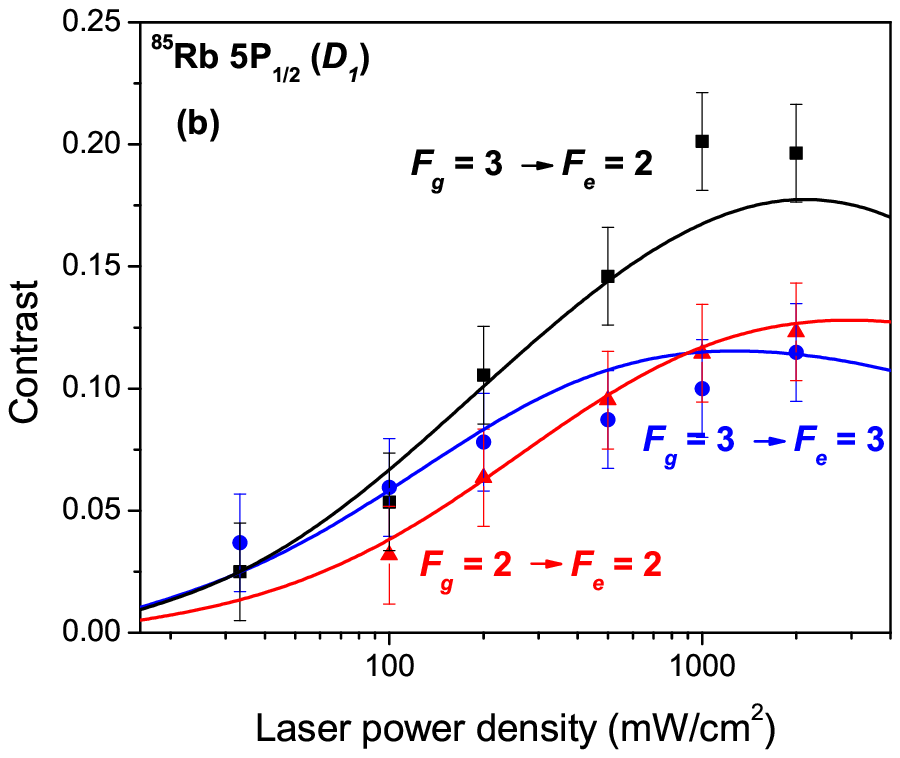}}
    \caption{\label{fig:contrast} (color online) Resonance contrast as a function of laser 
power density for (a) $^{87}$Rb and (b) $^{85}$Rb dark resonances. 
Markers, experiment; solid line, theory. $L=\lambda$.}
\end{figure*}

Among the parameters that one must know in order to fit the resonance shapes is the  Rabi
frequency $\Omega_R$, which is related to the laser power density $I$ as
$I=k_{Rabi}\Omega_R^2$.  
Figure~\ref{fig:rb87_11_rabi} shows the
$F_g=2\rightarrow F_e=1$ transition of $^{87}$Rb at various laser
power densities. The filled squares show the results of the
experiment, whereas the solid line shows the result of the
theoretical calculation. Figure~\ref{fig:contrast} shows the
measured and calculated resonance contrasts for all dark 
resonances of the rubidium $D_1$ transition. Similar to the case of
the ordinary cell, the resonance contrast and width increased with
increasing Rabi frequency, but the increase reaches a maximum at higher
Rabi frequencies and then diminishes.

The model assumed a residual Doppler width in the direction
transverse to the ETC walls on order of 60 MHz (FWHM) for all
calculations corresponding to wall separation $L=\lambda$. This Doppler width
corresponds to the value obtained by fitting a Voigt profile to the LIF 
excitation spectrum for $L=\lambda$ (similar to Fig.~\ref{fig:LIF_excitation}, 
except that in Fig.~\ref{fig:LIF_excitation}, $L=\lambda/2$.) The Doppler width
in the direction transverse to the ETC walls was therefore about ten times
less than the Doppler width in the directions parallel to the walls at the ETC
temperature. Such a Doppler width is consistent with the value of the
ground state relaxation rate $\gamma_g$ assumed to achieve the best fit with the
experimental data, if one assumes that the ground state relaxation rate is related
to the typical time of flight before an atom collides with one of the ETC walls. 

\begin{figure*}[htbp]
    \centering
         \resizebox{\columnwidth}{!}{\includegraphics{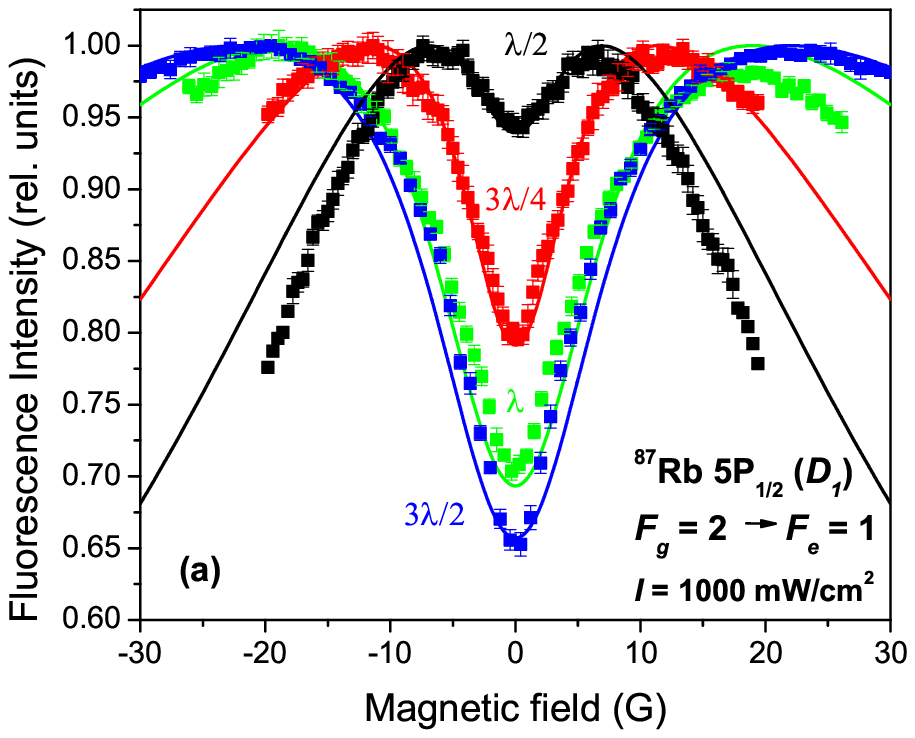}} 
         \resizebox{\columnwidth}{!}{\includegraphics{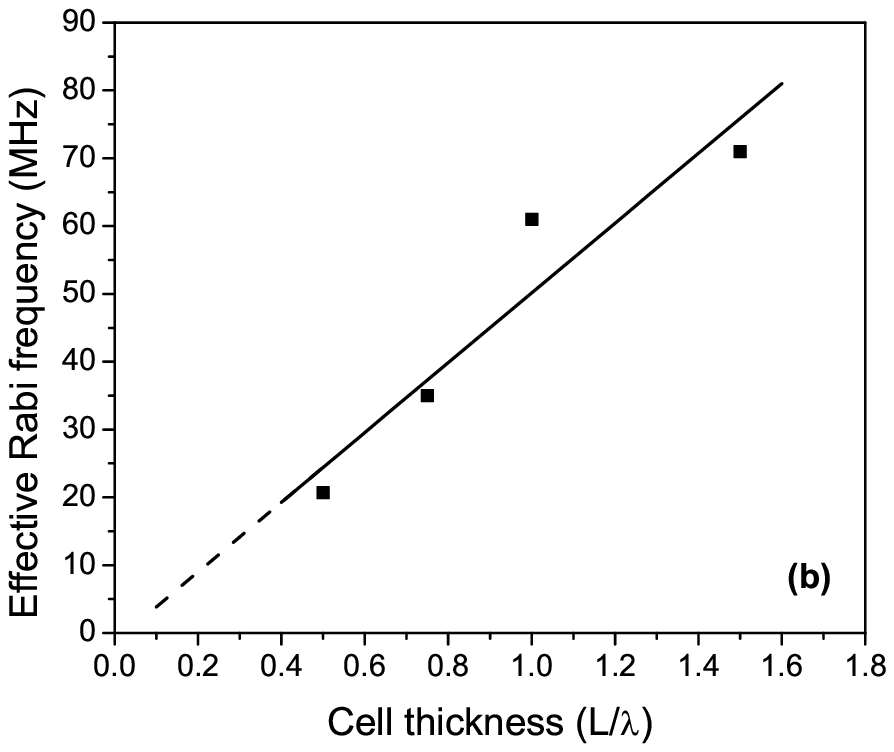}}
    \caption{\label{fig:etc_width} (color online) (a) Fluorescence intensity versus magnetic 
field at various values of the wall separation for the $F_g=2\rightarrow F_e=1$ 
transition of $^{87}$Rb. Markers, experiment; solid line, theory. (b) Effective Rabi frequency used in model to 
calculate (a) versus cell thickness.}
\end{figure*}

Since the ETC cell walls are inclined toward each other, it is possible to perform the
experiment at different wall separations. 
The approximate wall separation $L$ can be inferred from the constructive
($L=(2n+1)\lambda/4$) or destructive ($L=n\lambda/2$) interference
of the laser beams of wavelength $\lambda$ reflected from the
ETC's two inside walls. Figure~\ref{fig:etc_width} shows the laser
induced fluorescence intensity versus the magnetic field for
$L=\lambda/2$, $L=3\lambda/4$, $L=\lambda$, and $L=3\lambda/2$. The
theoretical curves are calculated with the residual Doppler width,
the ground state relaxation time $\gamma_g^{-1}$, and the excited state relaxation time $\gamma_e^{-1}$ 
all linearly proportional to the wall separation, which seems reasonable in the case of the ETC. 
The LIF-induced background was assumed to be 67\% for
$L=\lambda/2$, 57\% for $L=3\lambda/4$, and 50\% for $L=\lambda$ and $L=3\lambda/2$. 
The calculated curves follow the trend of the experimentally measured curves. However, 
in order to obtain a satisfactory fit, it was necessary to assume that the Rabi frequency is
modified by the small wall separation. Thus, an effective Rabi frequency was also adjusted
until the best fit could be obtained Interestingly, 
Fig.~\ref{fig:etc_width} suggests that the effective Rabi frequency used in the fit
is linearly proportional to the cell thickness, and the fit of the Rabi frequencies versus
wall thickness passes very close to the origin. 

Other parameters included were 10 MHz for the laser linewidth as
well as the atomic parameters from~\cite{Steck:rubidium87,Steck:rubidium85}. 

In general the theoretical calculations did a reasonably good job in
reproducing the resonance contrasts over a wide range of different
experimental conditions and for four distinct hyperfine transitions
in each of two isotopes. However, the agreement achieved between theory and
experiment in the ETC was somewhat less satisfactory than in the case of the ordinary
cell~\cite{Auzinsh:2009}, because in the ETC it was not possible to use the same
background and conversion between Rabi frequency and laser power density for all transitions. 
At high Rabi frequencies, the model
also ceases to work as well, just as was the case for the ordinary cells (see Fig. 9 in \cite{Auzinsh:2009}).
Furthermore, the agreement at large fields
 is also not perfect for all transitions. The inclusion
in the model of mixing of magnetic sublevels in the magnetic field turns out to be a small effect that
cannot significantly improve the large field results. For example, even at a magnetic field of 60 G, 
magnetic field mixing of the excited state of $^{85}$Rb changes the signals by only 0.5\%. 

There are several factors that could explain discrepancies between theory and experiment in the ETC. 
For one thing, the resonances in the ETC are very
sensitive to laser detuning (see Fig.~\ref{fig:detuning}).
Thus, the inevitable small drift of the laser
frequency during a single measurement or between measurements, which
had negligible impact on the results in the ordinary cell, had a greater 
impact on the results of the ETC. Also, the ETC cell thickness is a 
rapidly changing and nonlinear function of position. The
neutral density filters used to obtain different laser power
densities could displace the laser beam slightly to an area
with a different cell thickness and hence 
relaxation times. Furthermore, the laser beam was about 0.4
mm in diameter, and so the ETC thickness was not constant within the
laser beam. On the other hand,
the model was able to describe the data in the ETC at higher laser
power densities, because in the ETC, which has a larger ground state
relaxation rate, the saturation
parameter $\Omega_R^2/(\Gamma \gamma_g)$ is smaller than in the
ordinary cell for the same $\Omega_R^2$.

It should be noted that the experimental dependence on
the frequency detuning $\Delta$ of the 
width and contrast of the dark resonances formed in the ETC 
(Fig.~\ref{fig:detuning}) were in agreement with the 
results of the dark resonance formation in an ETC in a $\Lambda$-system 
with two lasers ($^{85}$Rb 5S$_{1/2}$,$F_g=2\rightarrow$ 5P$_{3/2}\rightarrow$ 
5S$_{1/2}$,$F_g=3$)~\cite{Sargsyan:2006,Pashayan:2007}. 
Namely, as the coupling laser was detuned from the resonance with an 
atomic transition, a strong increase of the resonance width 
of the electromagnetically induced transparency  
and a worsening of the contrast were recorded. In ordinary cm--size
cells, effectively the opposite behavior would be observable.

When the coupling laser frequency was in exact resonance with 
corresponding atomic transition in the above mentioned case of 
the formation of dark resonances in a $\Lambda$-system in an ETC, a weak dependence 
of the dark resonance width on $L$ (as $\sim 1/L^{1/4}$) was 
observed~\cite{Sarkisyan:2009}. The dependence of linewidth on $L$ presented 
in Fig.~\ref{fig:etc_width} was somewhat stronger. This could have been caused 
by a rapid increase of the width of the dark resonance ($\sim 1/L$) 
as the thickness decreased for the case of large $\Delta$~\cite{Sarkisyan:2009}.

\section{\label{Conclusion:level1}Conclusion}
Nonlinear magneto-optical resonances have been measured for all
hyperfine transitions of the $D_1$ line of $^{85}$Rb and
$^{87}$Rb in an ETC under a wide variety of experimental conditions,
which included different laser power densities, laser detunings, 
and ETC thicknesses. Dark resonances have been
observed as expected for all hyperfine transitions with $F_g\geq
F_e$. Bright resonances were not observed for the transitions with
$F_g<F_e$, which was consistent with the theoretical predictions. 
One of the main differences between these
resonances in the ETC and the resonances in an ordinary vapor cell
was that resonances in an ETC were substantially broader and had
smaller contrast. The experimental signals were described with a
theoretical model that was based on the optical Bloch equations, and which 
had been used previously to describe successfully these types
of resonances in ordinary vapor cells. Compared to more complex models 
(see, for example, \cite{Andreeva:2007b}), our model
achieved better agreement when values for the
model parameters were used that corresponded to the particular
characteristics of the ETC. 
The theoretical model suggested that the ground state relaxation rate $\gamma_g$ was
a key parameter for describing the signals, but that its value in the
ETC depended on the ETC wall separation instead of on the laser beam
diameter, as in ordinary vapor cells. Furthermore, in order to obtain
the best possible agreement between theory and experiment, it was necessary
to introduce an additional excited state relaxation rate $\gamma_e$ that was related to the ETC
wall separation, and which was negligible in the model for the ordinary cell. It was also
necessary to introduce an effective Rabi frequency that depended on wall separation. 
The best-fit parameters indicated that the atoms in the ETC had a residual Doppler
distribution in the direction perpendicular to the cell walls that
was on the order of 60 MHz (FWHM), which was consistent with a ground state relaxation
rate based on wall collisions.
 
The results showed that it is
possible to describe some aspects of nonlinear magneto-optical resonances in
ordinary cells and ETCs over a wide range of experimental conditions
with the same theoretical model, which was based on the optical Bloch
equations, and averaged over the Doppler profile, included all
neighboring hyperfine transitions, took into account the splitting and mixing of
the magnetic sublevels in an external magnetic field, and 
treated the coherence properties of the radiation field. As a result,
this theoretical model can serve as a tool for future
investigations with ETCs and for the development of practical
applications based on them. It should be noted that the conditions of the ETC may not be stationary as in the
case of ordinary cells. Our attempt to account for non-stationary behavior by introducing
an effective Rabi frequency that depends on the wall separation probably could be 
improved in the future by a more detailed model. The remaining discrepancies 
between theory and experiment suggest that there do exist
additional physical effects that should be taken into account or treated in 
greater detail in the theoretical description of the ETC.

\begin{acknowledgments}
The Riga group would like to thank Maris Tamanis for assistance with the experiments 
and Christina Andreeva for useful discussions and to acknowledge  
support from the Latvian National Research Programme in Material Sciences 
Grant No. 1-23/50 and the Latvian Science Council Grant No. LZP 09.1196. 
The work in Ashtarak was supported in part by the INTAS South-Caucasus Grant
06-1000017-9001. L.~K. acknowledges support from the ESF project Nr. 2009/0138/1DP/1.1.2.1.2./09/IPIA/VIAA/004, 
and F.~G. acknowledges support from the ESF project Nr. 2009/0223/1DP/1.1.1.2.0./09/APIA/VIAA/008. 
\end{acknowledgments}
\bibliography{rubidium}
\end{document}